# 360° Domain Walls: Stability, Magnetic Field and Electric Current Effects

**Jinshuo Zhang[1], Saima A. Siddiqui[2], Pin Ho[1], Jean Anne Currivan-Incorvia[2,3], Larysa Tryputen[1], Enno Lage[1], David C. Bono[1], Marc A. Baldo[2] and Caroline A. Ross[1]**

1. Department of Materials Science and Engineering, Massachusetts Institute of Technology, Cambridge MA, 02139
2. Department of Electrical Engineering and Computer Science, Massachusetts Institute of Technology, Cambridge MA, 02139
3. Department of Physics, Harvard University, Cambridge MA, 02138

**Abstract:** The formation of 360° magnetic domain walls (360DWs) in Co and $Ni_{80}Fe_{20}$ thin film wires was demonstrated experimentally for different wire widths, by successively injecting two 180° domain walls (180DWs) into the wire. For narrow wires ($\leq$ 50 nm wide for Co), edge roughness prevented the combination of the 180DWs into a 360DW, and for wide wires (200 nm for Co) the 360DW collapsed, but over an intermediate range of wire widths, reproducible 360DW formation occurred. The annihilation and dissociation of 360DWs was demonstrated by applying a magnetic field parallel to the wire, showing that annihilation fields were several times higher than dissociation fields in agreement with micromagnetic modeling. The annihilation of a 360DW by current pulsing was demonstrated.

Magnetic domain walls (DWs) in narrow wires provide a data token for devices such as racetrack memory and logic gates [1-3]. DW devices maintain the traditional merits of magnetic data storage including non-volatility and high density, but offer new functionality including fully electrical operation by using spin torque transfer to manipulate the DWs and magneto-resistance to detect them. This can in principle enable faster switching speeds and lower energy consumption compared to DRAMs and other semiconductor devices [4]. Proposed DW devices are made from nanowires containing head-to-head 180DWs, where the orientation of magnetization rotates through 180° [5]. The 180DW in a wire with in-plane magnetization adopts a vortex or a transverse configuration, with transverse DWs favored in narrow wires [6] as shown in the upper panel of Fig. 1(a). An essential requirement for DW devices is to be able to translate DWs within the device, which can be accomplished using a magnetic field or a current pulse due to spin transfer torque [7, 8].

Closely spaced 180DWs in a nanowire interact magnetostatically, and the attraction between them can lead to the annihilation of a pair of DWs, or the formation of a metastable 360° domain wall (360DW) when two 180DWs of opposite sense combine, as shown in the lower panel of Fig. 1(a). 360DWs are also known as 1-D skyrmions [9], an example of a class of topologically protected structures which are of intensive study due to their stability and low critical driving field [10]. 360DWs have been observed both in continuous ferromagnetic films [11] and in ferromagnetic nanostructures such as thin film rings or ellipses [12-14]. The orientation of magnetization rotates through 360°, and due to the opposite sense of core magnetization in the two component 180DWs, magnetic flux closure reduces the stray field around the 360DW compared to that of a 180DW [15]. 360DWs are not expected to be translated by an applied field, but instead can be dissociated or annihilated. However, micromagnetic simulations predict that a current can translate the 360DW via spin torque transfer [15]. Moreover, instead of Walker Breakdown as observed in 180DWs, a 360DW is predicted to undergo annihilation at a sufficiently high spin current density. Simulations predict that 360DWs of different chirality can be filtered [16], and 360DWs have been proposed for use in magnetic sensors [17] and as an alternative to 180DWs in memory and logic.

360DWs in magnetic wires have been detected using magnetic force microscopy (MFM) [11], scanning electron microscopy with polarization analysis [14], and anisotropic magnetoresistance (AMR) measurements [18-21], in which the formation of a 360DW results in a decrease of resistance. There has been plenty of theoretical and modeling work on the behavior of 360DWs [22-24], however, there have been no systematic experimental reports on the formation of 360DWs, their response to a field as a function of wire geometry, nor any observation of current-driven motion of a 360DW. In this article, we first demonstrate the formation and stability of 360DWs in specifically designed Co and NiFe nanostructures of different widths with in plane magnetization. We then demonstrate the effect of an applied magnetic field to dissociate or annihilate the 360DWs using both AMR and MFM measurements, and relate the results to micromagnetic simulations.

The method used to generate a 360DW is similar to previous work [14]. The structures consist of a magnetic thin film wire in the shape of an arc, with width varying from 50 nm to 200 nm, connected to a round injection pad of 1 $\mu$m diameter. The structures were made from a thin film stack of Ta (5nm)/ $Ni_{80}Fe_{20}$ (10nm)/ Au (5nm) or Ta (5nm)/ Co (10nm)/ Au (5nm) which was deposited by magnetron sputtering (5 cm diameter target, 100 W, and growth rates of 0.15-0.30 nm s$^{-1}$ for different materials) on a Si substrate with native oxide, at an Ar pressure of 1 mTorr and a base pressure better than $5 \times 10^{-8}$ Torr. A bilayer resist [25] consisting of 4% polymethyl-methacrylate (PMMA) of thickness 30 nm and 2% hydrogen silsesquioxane (HSQ) of thickness 40 nm was spin coated on the magnetic film then exposed using an Elionix F-125 e-beam lithography tool with a dose of 38 mC cm$^{-2}$. The HSQ layer was developed using 4% NaCl + 1% NaOH in water solution for 20 seconds, washed with DI water and carefully dried by nitrogen blow guns. Then the underlying PMMA layer was removed with $O_2$ plasma at a pressure of $6 \times 10^{-3}$ Torr with 90 W power for 2 minutes. Using these patterned bilayer resists as an etch mask, the metal film was then etched using an Ar ion beam etch with a beam current of 10 mA at a pressure $2 \times 10^{-4}$ Torr. The etching was monitored using an end-point detector. Since the PMMA acts as a sacrificial layer, the resist stack could then be removed with hot 1165 solvent [*MicroChem* Corp.] and sonication after the ion beam etching. An AFM image of a typical structure is shown in the upper panel of Fig. 1(b).

In order to generate a 360DW, an in-plane field sequence perpendicular to the arc was applied. $H_y = +3000$ Oe was applied to fully saturate the magnetization and form the first 180DW with its core magnetized along +y at the center of the arc at remanence. $H_y = -300$ Oe was then applied, a field sufficient to reverse the magnetization in the round pad but not high enough to reverse the magnetization in the arc due to its higher shape anisotropy. A second 180DW with opposite sense to the first 180DW was formed at the interface between the round pad and the arc. The two 180DWs combined to form a 360DW [14]. AFM and MFM images of a Co sample of width 120 nm are shown in the lower image of Fig. 1(b). The dark and bright contrast observed at the center of the arc confirmed the presence of a 360DW.

Multiple Co samples with different arc widths of 50 nm, 80 nm, 120 nm, 150 nm and 200 nm were fabricated to study the 360DW stability as a function of arc width. We expect TWs to be energetically preferable within the width range according to micromagnetic simulations in Co nanowires. The 80 nm, 120 nm and 150 nm samples successfully formed a 360DW in the arc. In the sample with 50 nm wire width, two 180DWs of opposite sense were formed but remained separate in the arc without combining into a 360DW (the bright contrast originates from the second 180DW). This is attributed to pinning due to edge roughness: the amplitude of line edge roughness is expected to be independent of linewidth [25], but the resulting changes in linewidth are proportionately larger for narrower wires and lead to stronger pinning. The length of a 180DW also decreases with linewidth [2], making them more sensitive to high frequency edge roughness. As a result, the extrinsic pinning in the 50 nm wide wire is believed to explain why the formation of a 360DW was prevented.

For the sample with wire width of 200 nm, no 360DW was observed. MFM images of the sample after applying the initial saturation field of $H_y = +3000$ Oe showed a 180DW with transverse configuration as illustrated by dark contrast at the arc center. This result shows that a 180DW could be formed but the second 180DW is assumed to annihilate the first one instead of forming a 360DW. This is attributed to the reduced stability of a

360DW in a wider wire. Moreover, the weaker edge roughness pinning may have enabled the second 180DW to approach the first at a higher velocity, which promotes annihilation.

The reduced stability of a 360DW with increasing wire width in Co samples was demonstrated by calculating the critical fields for annihilation and dissociation of a 360DW with dependence of wire width. Fig. 2 shows micromagnetic simulations using OOMMF [26] of the effect of a field $H_x$ applied along the wire length, using parameters similar to those of $Ni_{80}Fe_{20}$. The exchange constant was $A = 10^{-5}$ erg cm$^{-1}$, saturation magnetization $M_s = 1000$ emu cm$^{-3}$, anisotropy factor $K_u = 0$, $\alpha = 0.02$ and $\beta = 0.03$. The cell size was (5 nm)$^3$, and the length and thickness of the simulated wires were 10 $\mu$m and 10 nm, respectively. We first initiated a 360DW at the center of the wire, allowed it to relax, then applied a field in either the +x or –x direction along the wire to dissociate or annihilate the 360DW. We increased the field from 0 with a step size of 10 Oe in order to find the critical value. As shown in Fig. 2, the critical fields for annihilation and dissociation decreased by about 40% and 80%, respectively with an increase of wire width from 50 nm to 300 nm. The 360DW became less symmetrical as the width increased, with the two component 180DWs tilting towards each other at one side of the wire as shown in te insets of Fig. 2, making the 360DW less stable. It has also been shown in Ref. [18] that the 360DW forms a much less stable vortex structure with increasing wire width.

The effect of magnetic field on 360DWs was studied experimentally in both $Ni_{80}Fe_{20}$ and Co samples and characterized by AMR and MFM, respectively. We used AMR measurements instead of MFM for $Ni_{80}Fe_{20}$ samples because the DWs in $Ni_{80}Fe_{20}$ samples are easily perturbed by the field from the MFM tips. AMR measurements have the advantage of enabling in situ measurement, at the cost of more complex sample preparation requiring 4-point electrodes which were made of Ta (7nm)/Au (100 nm) and patterned using liftoff over the pad-wire structure in a second lithography step. A scanning electron microscopy (SEM) image of a $Ni_{80}Fe_{20}$ sample is shown in Fig. 3(a), which was 200 nm wide and 10 nm thick. Vortex DWs are energetically favorable in this geometry from simulations. A small DC current was applied between the outer two electrodes and the voltage between the inner two electrodes was measured in order to obtain the resistance. This reference DC current was lower than 5 $\mu$A, corresponding to an average current density of $< 2 \times 10^9$ A/m$^2$ in the arc, which was 0.001 – 0.01 times the current density reported to move a DW in $Ni_{80}Fe_{20}$ nanowires [8, 27-30]. Therefore the effect of the measurement current on the wall was neglected.

As shown in the round dot line in Fig. 3(b), a head-to head 180DW was formed by applying $H_y = +3000$ Oe, then a field in the +x direction was applied, starting from zero and increasing with a step size of 1 Oe. The resistance was measured after each field step. A change of $\Delta R = 0.05$ Ω was observed at 5 Oe, indicating movement of the 180DW to the right, out of the area between the inner two electrodes. The increase of the resistance is due to the AMR effect from the 180DW: the resistance is lower at the domain wall because the magnetization is locally perpendicular to the electron flow. The AMR follows the relation $\frac{\Delta \rho(H)}{\rho_{av}} = \frac{\Delta \rho}{\rho_{av}} \left( \cos^2 \theta - \frac{1}{3} \right)$ (1), in which $\Delta \rho(H)$ is the change of resistivity and $\theta$ is the angle between magnetization and electron flow [31]. If a field was applied in the –x direction, a similar resistance jump of $\Delta R = 0.05$ Ω was observed at -6 Oe as shown in the square dot line in Fig. 3(b), indicating the 180DW moving to the left, out of the area between the inner two electrodes.

Each test was repeated 10 times and the same resistance jump was observed 8 and 9 times for fields along +x and -x, respectively. The other tests gave either no resistance change or a change of 0.03~0.04 Ω indicating differences in the domain wall structure. Other groups have also shown domain wall AMR in large numbers of repeated tests [18].

A field sequence of $H_y = +3000$ Oe followed by $H_y = -300$ Oe was then used to form a 360DW in the 200 nm wide $Ni_{80}Fe_{20}$ wire, followed by a field in the +x or –x direction as shown in Fig. 3(c) and (d), respectively. In Fig. 3(c), a resistance jump of $\Delta R \approx 0.07$ Ω was observed at $H_x = +14$ Oe. The higher resistance change confirmed the presence of a 360DW instead of 180DW and the higher field of $H_x = +14$ Oe represents the critical field to dissociate a 360DW, overcoming the magnetostatic attraction between the two component 180DWs and separating them to form a reverse domain. In comparison, in Fig. 3(d), a resistance jump of $\Delta R \approx 0.07$ Ω was observed at $H_x = -84$ Oe, much higher than in Fig. 3(c). This represents the critical field to annihilate the 360DW, in which the field compresses the 360DW, eventually eliminating it.

From Fig. 2, for a model nanowire of width $w$ = 200 nm and thickness $t$ = 10 nm, the critical field to dissociate a 360DW is $H_x = +75$ Oe and to annihilate it is $H_x = -310$ Oe. The modeling predicted much higher absolute values of the annihilation and dissociation fields than were measured experimentally, which may be a result of the zero temperature of the model and the lack of edge roughness and other defects that could initiate annihilation and dissociation. However, both model and experiment agree in showing annihilation field several times larger in magnitude than dissociation fields.

The expected ratio of AMR between a 180DW and a 360DW was calculated by exporting the magnetization distribution from OOMMF into MATLAB and determining the resistance based on equation (1). The AMR ratio was $\frac{\Delta R_{360}}{\Delta R_{180}} = 1.46$ for a $w = 200$ nm and $t = 10$ nm Ni$_{80}$Fe$_{20}$ nanowire, similar to the experimental result of $\frac{\Delta R_{360}}{\Delta R_{180}} \approx \frac{0.07\ \Omega}{0.05\ \Omega} = 1.4$. The agreement is excellent considering the error in the resistance measurement of about $\pm 0.005\ \Omega$ estimated from Fig. 3(b)-(d).

The formation and field effects on 360DWs in Co nanowires were measured using MFM. As shown in Fig. 4(a), a 360DW was first formed in a Co sample with wire width of 80 nm by applying $H_y = +3000$ Oe then $H_y = -300$ Oe. Different magnetic fields were applied in the +x direction to dissociate the 360DW. The 360DW was unchanged for $H_x \leq +100$ Oe, but after the application of $H_x = +150$ Oe, the two component 180DWs moved apart forming a reverse domain. After the application of $H_x = +200$ Oe, the component 180DWs moved further such that the left-hand bright-contrast wall moved into the injection pad and the right dark-contrast wall moved to the end of the wire. The reversal of the wire is evident from the change in contrast from bright to dark at the tip of the wire, labeled by the black circles in Fig. 4. A similar dissociation process was observed in a Co sample with 150 nm wire width, Fig. 4(b), except that the field steps did not capture the presence of two separate 180DWs in the wire. In the 150 nm wide wire, dissociation occurred at a field between 125 Oe and 150 Oe.

360DW annihilation in a Co sample was demonstrated by applying a field in the $-x$ direction. The same field sequence as in Fig. 4 was applied initially to form a 360DW at the center of the arc. The 360DW remained stable in the arc for $-H_x \leq 500$ Oe. The 360DW disappeared after applying a field of $H_x = -700$ Oe, but different from the dissociation results in Fig. 4, the right-hand end of the wire retained its bright contrast before and after the annihilation. This proves that the magnetization direction remained the same in the arc, indicating annihilation instead of dissociation of the 360DW. The field for annihilation (-700 Oe) was also much larger than that for dissociation (+150 Oe) in the 150 nm wide Co wire, confirming the same trend as seen in the Ni$_{80}$Fe$_{20}$ samples and in micromagnetic simulations.

The effects of current on 360DWs were examined in 150 nm wide Co samples. Two electrodes were placed on top of the sample as shown in Fig. 5 (a), labeled I$^+$ and I$^-$, and the conventional current flowed from the right electrode to the left electrode. Fig. 5(b) is an MFM image of the sample initiated with a 360DW at the arc center. Current pulses of +2 V, +4 V or +6 V amplitude and 200 ns duration were then injected into the sample, corresponding to current densities of approximately 0.8, 1.6 or 2.4×10$^{12}$ A/m$^2$ respectively in the arc. A delay of 3 seconds was inserted between each pulse to minimize Joule heating [32]. From MFM imaging there was no observed effect of current pulses with amplitude of 2 V. Fig. 5(c) shows the wall after 2 pulses of 4 V, which moved ~200 nm to the right, indicating a small current-driven 360DW motion. We did not see any further translation after 2 additional pulses of 4 V. After 10 additional identical pulses of 4 V the 360DW disappeared as shown in Fig. 5(d), which indicates that the 360DW was annihilated.

Previous simulation results [15] predicted that a spin current will translate a 360DW, and at high enough spin current density will annihilate it. However, the experiment showed annihilation after only a small translation. The limited current-driven motion of the 360DW is attributed to extrinsic pinning from edge roughness. This was confirmed by measurements of current-driven motion of 180DWs in the same sample. Ten identical 200 ns current pulses were injected at voltages of +2 V to +6 V (the highest voltage led to damage to the sample), but the 180DW was not translated. This suggests that DWs are strongly pinned in the Co nanowires [33], and the current therefore annihilated the 360DW without moving it.

The current pulses provided an effective method for annihilating the 360DW. As shown above, it required -700 Oe to annihilate a 360DW in the 80 nm wide Co sample but only +150 Oe to move a 180DW. For the current pulsing, annihilation of the 360DW occurred at 4 V pulse amplitude but moving a 180DW required at least 6 V. The joule heating during current pulsing may contribute to destabilizing the wall, but the temperature decreased too quickly to enable measurement [32]. We estimate that the temperature increase would be ~10 K based on the results in ref. [34], which is also consistent with the measurement based on resistance change [35]. The modeling illustrates the different 360DW annihilation mechanism between using current pulses and using external field, making current pulses particularly effective in 360DW annihilation.

Figs. 6(a) and 6(b) show snapshots of simulated 360DW annihilation process by field and current in a 100 nm wide and 10 nm thick wire, using the same parameters as in Fig. 2. In Fig. 6 the 360DW develops a trapezoidal shape with the wider side at the bottom of the figure. In the field-driven annihilation process, the field pushes the two component 180DWs together and the 360DW is annihilated starting from its wider edge. On the contrary, in the current-driven annihilation process, the current translates the 360DW, which moves with an oscillatory motion, eventually closing up from the narrower edge. The current- and field-driven annihilation can be better understood by comparing the maximum exchange energy of the 360DW during the annihilation. At the critical annihilation field of -360 Oe, the maximum exchange energy was $1.09 \times 10^{-17}$ J, while at the critical annihilation spin current velocity of $u = 200$ m/s (corresponding to a current density of $J = 7.0 \times 10^{12}$ A/m$^2$), the maximum exchange energy was only $0.71 \times 10^{-17}$ J, meaning that the current-driven annihilation needs to overcome a lower exchange energy barrier than the field-driven annihilation.

In summary, 360DWs were formed in both $Ni_{80}Fe_{20}$ and Co wires with a range of widths. 360DW dissociation and annihilation was experimentally observed by AMR measurements in $Ni_{80}Fe_{20}$ and by MFM measurements in Co nanowires, and the critical fields were determined. The experimental results were compared with micromagnetic simulations, both of which showed that field-driven annihilation requires several times larger field than field-driven dissociation, though the magnitude of the critical fields were higher in the simulations. The effects of current on 360DWs in Co were examined using MFM. Current produced a small translation of the 360DW then led to its annihilation at $1.6 \times 10^{12}$ A/m$^2$, a current smaller than that required to translate a 180DW in the same width Co nanowire. The high fields required for translation are attributed to edge pinning in the Co, but the relatively low current required for annihilation is consistent with the lower energy barrier for current-driven annihilation compared to field-driven annihilation predicted by micromagnetic modeling. The modeling also predicts a different annihilation mechanism for current and field. These results provide experimental evidence for the formation and manipulation of 360DWs, which are important in the study of memory and logic devices based on both 180DWs and 360DWs.

**Acknowledgements:** This work was supported by the NSF under award EECS 1101798. Shared facilities of the MIT NanoStructures Laboratory and Center for Materials Science and Engineering (NSF DMR1419807) were used.

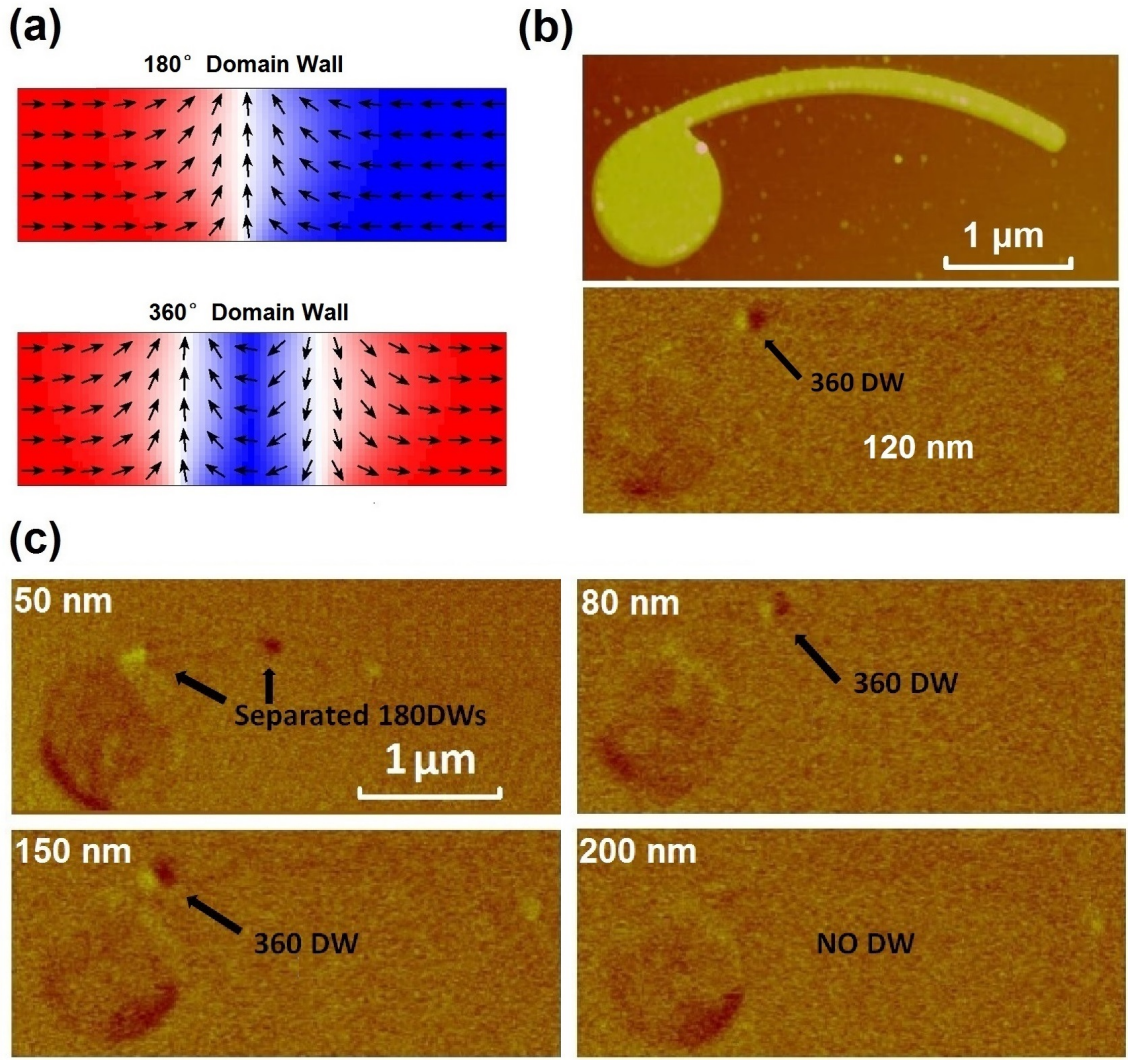

Figure 1. (a) Simulation of a transverse 180DW and a 360DW inside a Co wire, 10 nm thick and 120 nm wide. Arrows show magnetization, and red (positive) and blue (negative) indicate the component of magnetization along the wire length, the x-direction. (b) The upper AFM image shows a 120 nm wide Co wire and pad used to generate a 360DW, and the lower MFM image illustrates a 360DW in the wire; (c) MFM images of other Co samples with wire width of 50 nm, 80 nm, 150 nm and 200 nm.

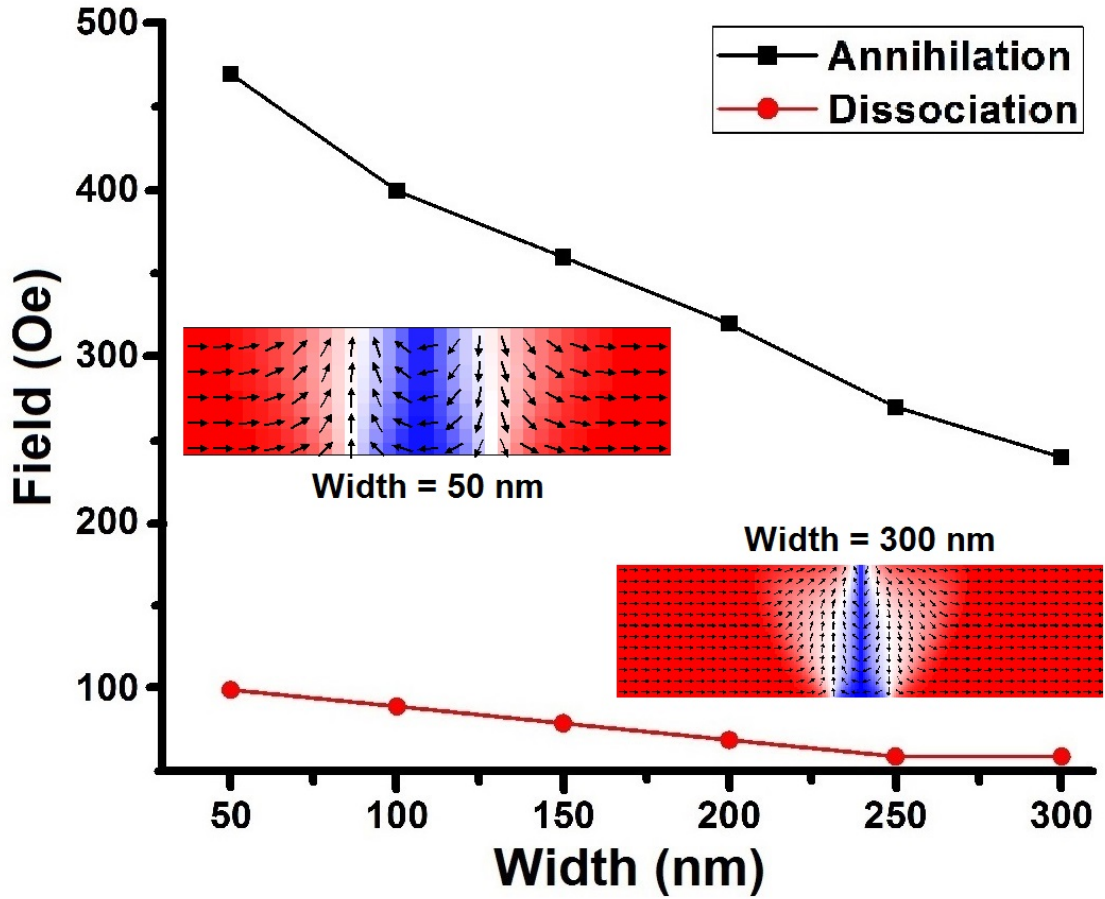

Figure 2. Calculation of the critical annihilation and dissociation fields vs. wire width in 10 nm thick wires, based on micromagnetic modeling of a wire with $M_s$ = 1000 emu cm$^{-3}$ and $K_u$ = 0. Images of the remanent 360DWs at 50 nm and 300 nm width are superposed.

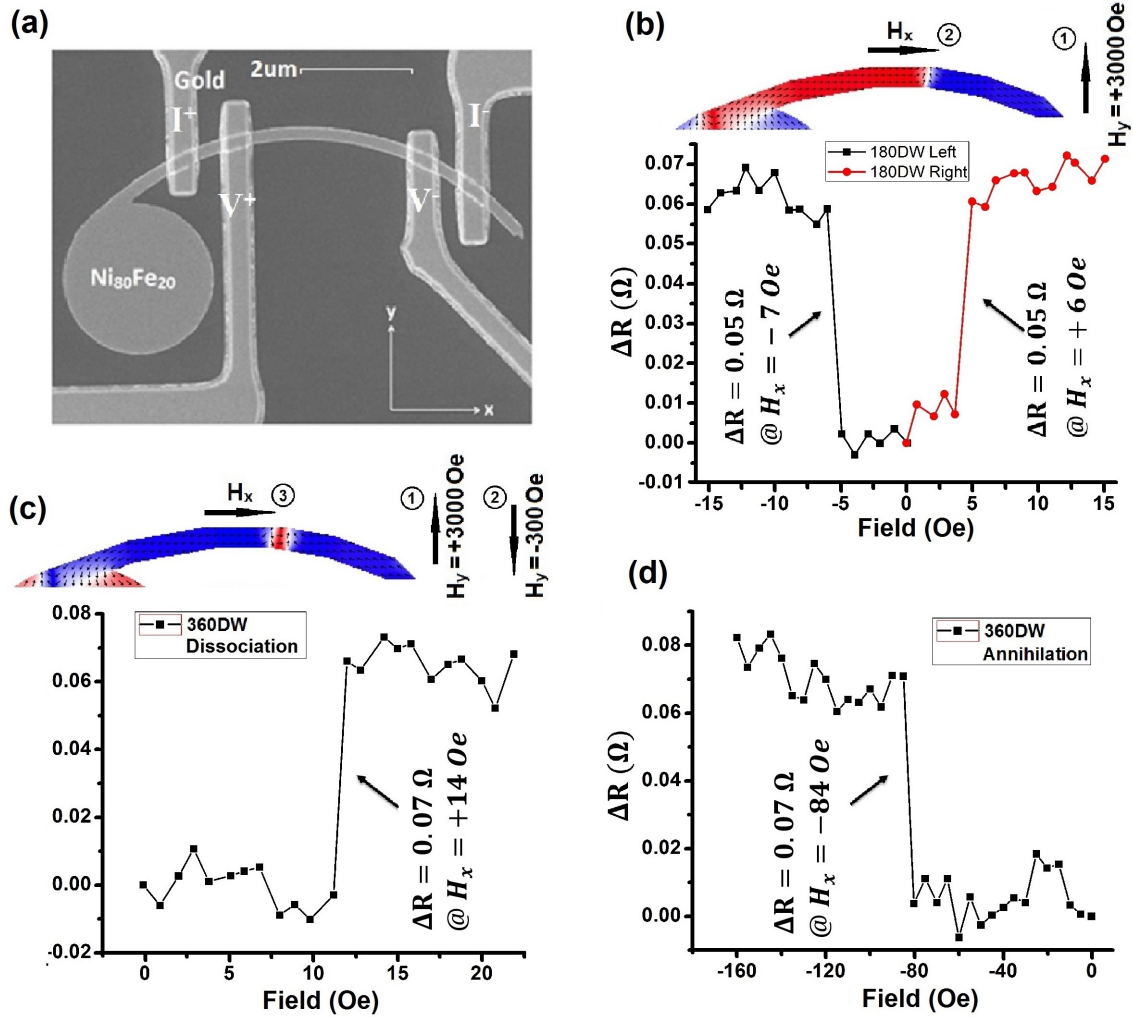

Figure 3. (a) SEM image of $Ni_{80}Fe_{20}$ sample with gold electrodes on top. A DC current less than 5 $\mu$A is injected from I$^+$ to I$^-$ to measure the AMR; (b) AMR measurements indicating a 180DW being moved left or right by a field $H_x$. The sequence of fields used to generate the wall then move it are shown with circled numbers, with negative $H_x$ moving the wall along the –x direction; (c) Schematic and model of the field sequence to form and then dissociate a 360DW. AMR measurements indicate dissociation in a field applied in the +x direction; (d) AMR measurements indicate annihilation of the 360DW in a field applied in the -x direction.

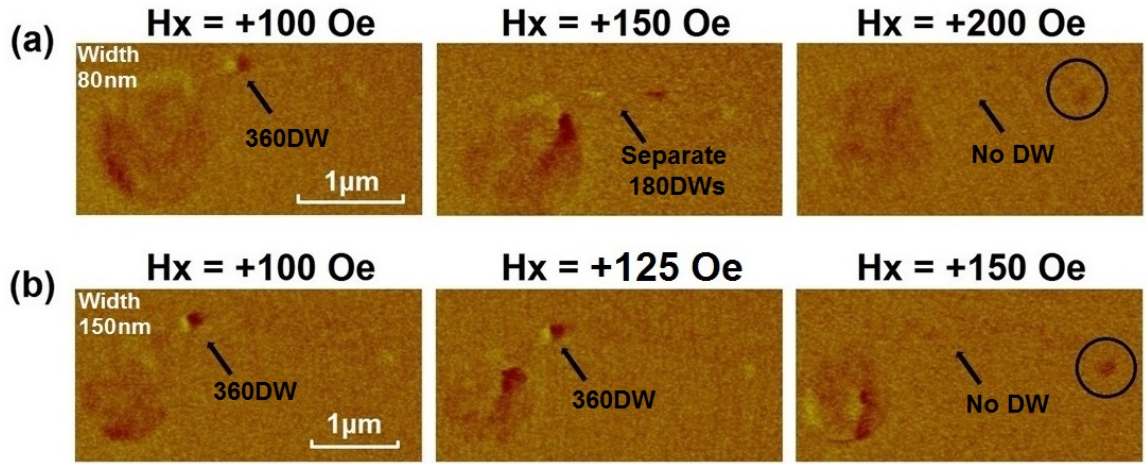

Figure 4. (a) A series of MFM images of a 360DW in a Co wire with width of 80 nm after application of different fields in the +x direction, showing separation into two 180DWs at +150 Oe and movement of the walls out of the wire at +200 Oe; (b) A series of MFM images of a 360DW in a Co wire with width of 150 nm after application of different fields in the +x direction. The wall was dissociated and the two 180DWs moved out of the wire between 125 Oe and 150 Oe. In both samples dissociation is evident from the change of contrast at the tip of the wire from light in the first panel to dark (black circle).

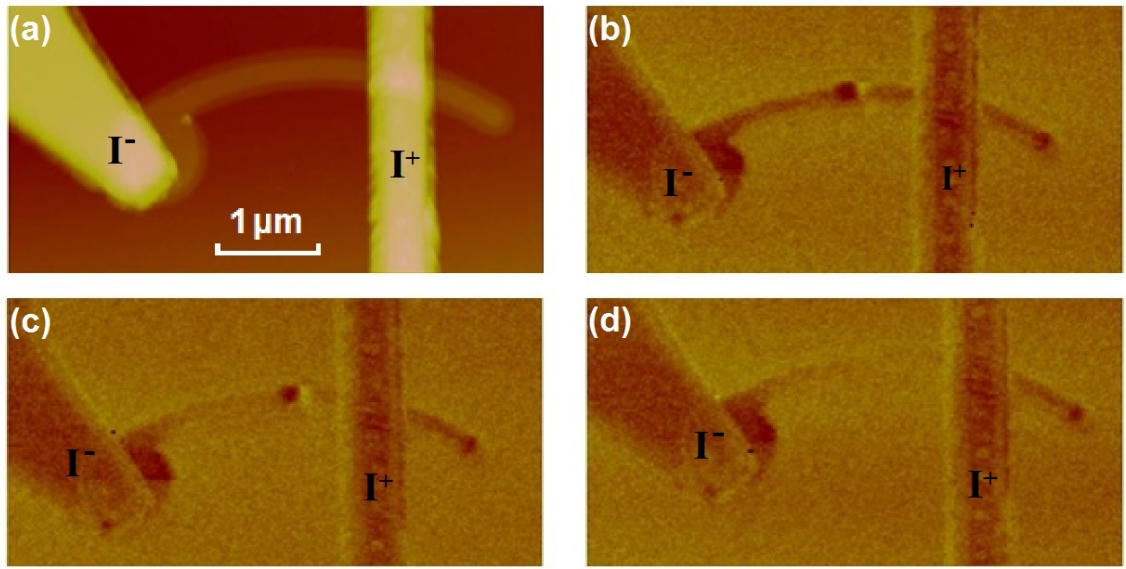

Figure 5. (a) AFM of Co sample with wire width 150 nm with gold electrodes on top, with I⁻ and I⁺ indicating the direction of current flow; (b) MFM of the sample after initializing a 360DW; (c) MFM of the sample after injecting 2 current pulses with +4V amplitude and 200 ns duration showing a small translation of the 360DW; (d) MFM of the sample after injecting 10 more current pulses with +4V amplitude and 200 ns duration, showing annihilation of the 360DW.

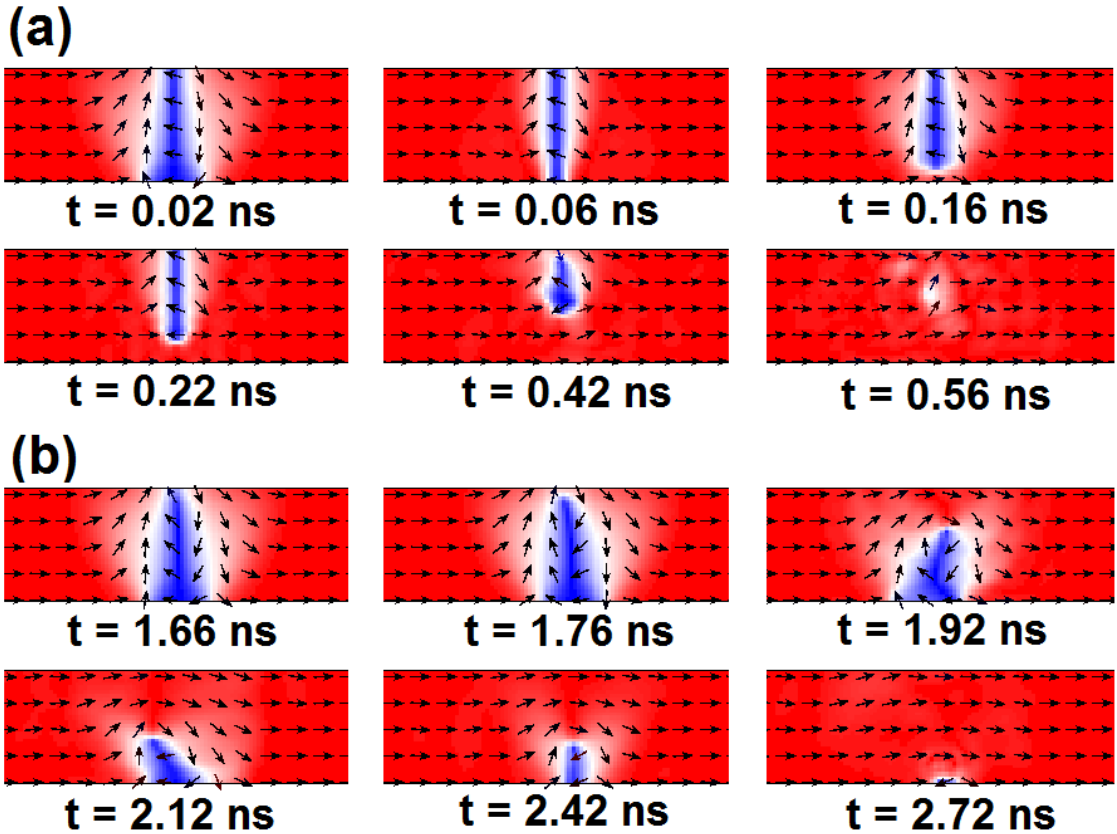

Figure 6. (a) Snapshots of OOMMF simulation results (wire thickness 10 nm and width 100 nm) of the annihilation of a 360DW by a field of $H_x = +400$ Oe; (b) Snapshots of OOMMF simulation results of the annihilation of a 360DW by a spin current of $u_x = +200$ m/s.